\documentclass{article}
\usepackage{amsmath, amssymb}
\usepackage{geometry}

\usepackage{soul}
\usepackage{textcomp}
\usepackage{paralist}
\usepackage{multirow}
\usepackage{subfigure}
\usepackage{graphics, graphicx}
\usepackage{epsfig}

\usepackage[super]{natbib} 

\usepackage{float}           
\floatstyle{plaintop}            
\restylefloat{table}

\usepackage[colorlinks,linkcolor=black,anchorcolor=black,citecolor=black]{hyperref}

\usepackage{threeparttable}
\usepackage{tabularx}

\usepackage{cleveref}

\usepackage{hyperref}
\usepackage{xr-hyper}

\makeatletter

\newcommand*{\addFileDependency}[1]{
\typeout{(#1)}
\@addtofilelist{#1}
\IfFileExists{#1}{}{\typeout{No file #1.}}
}\makeatother

\newcommand*{\myexternaldocument}[1]{%
\externaldocument{#1}%
\addFileDependency{#1.tex}%
\addFileDependency{#1.aux}%
}

\myexternaldocument{supplement}

\newcommand\Tstrut{\rule{0pt}{2.6ex}}         
\newcommand\Bstrut{\rule[-1.3ex]{0pt}{0pt}}   

\newcommand\norm[1]{\lVert#1\rVert}

\providecommand{\keywords}[1]
{
	\small	
	\textbf{\textit{Keywords---}} #1
}

\title{GGL-PPI: Geometric Graph Learning to Predict Mutation-Induced Binding Free Energy Changes}

\author{Md Masud Rana$^1$ and Duc Duy Nguyen$^1$ \footnote{Address correspondences to Duc Duy Nguyen. E-mail: ducnguyen@uky.edu}\\
$^1$ Department of Mathematics,
University of Kentucky, KY 40506, USA\\
}

\begin{document}

\maketitle

\begin{abstract}
Protein-protein interactions (PPIs) are critical for various biological processes, and understanding their dynamics is essential for decoding molecular mechanisms and advancing fields such as cancer research and drug discovery. Mutations in PPIs can disrupt protein binding affinity and lead to functional changes and disease. Predicting the impact of mutations on binding affinity is valuable but experimentally challenging. Computational methods, including physics-based and machine learning-based approaches, have been developed to address this challenge. Machine learning-based methods, fueled by extensive PPI datasets such as Ab-Bind, PINT, SKEMPI, and others, have shown promise in predicting binding affinity changes. However, accurate predictions and generalization of these models across different datasets remain challenging. Geometric graph learning has emerged as a powerful approach, combining graph theory and machine learning, to capture structural features of biomolecules. We present GGL-PPI, a novel method that integrates geometric graph learning and machine learning to predict mutation-induced binding free energy changes. GGL-PPI leverages atom-level graph coloring and multi-scale weighted colored geometric subgraphs to extract informative features, demonstrating superior performance on three validation datasets, namely AB-Bind, SKEMPI 1.0, and SKEMPI 2.0 datasets. 
Evaluation on a blind test set highlights the unbiased predictions of GGL-PPI for both direct and reverse mutations. The findings underscore the potential of GGL-PPI in accurately predicting binding free energy changes, contributing to our understanding of PPIs and aiding drug design efforts.
\end{abstract}

\keywords{geometric graph, machine learning, protein-protein interactions, mutation, binding free energy changes}

\section{Introduction}
Protein-protein interactions (PPIs) play a fundamental role in numerous biological processes, including cell signaling, metabolic pathways, and immune responses \cite{chuderland2005protein, jubb2015flexibility, gonzalez2012chapter}. Understanding PPIs and their dynamics is crucial for unraveling the intricate mechanisms underlying these processes and holds significant implications for various fields, such as cancer research, drug discovery, and personalized medicine \cite{gonzalez2012chapter, geng2019finding}. The effects of mutations on PPIs have drawn substantial attention due to their potential impact on protein function and cellular behavior \cite{moretti2013community, brender2015predicting, zhang2020mutabind2, rodrigues2019mcsm}. Missense mutations, which involve single amino acid substitutions, can disrupt the binding affinity between proteins and their partners \cite{li2014predicting, nishi2013cancer}. Such alterations can lead to malfunctioning PPI networks, resulting in diseases, drug resistance, or other molecular disorders \cite{gao2015insights, david2012protein, engin2016structure, stein2019biophysical, portelli2018understanding, vedithi2018structural}. Therefore, accurate prediction of the impact of mutations on binding affinity holds significant importance in understanding disease mechanisms, facilitating therapeutic interventions, and enabling the design of innovative biopharmaceuticals.

One of the key parameters used to assess the impact of mutations on PPIs is the binding free energy change ($\Delta\Delta G$). This thermodynamic parameter quantifies the difference in binding affinity between the wild-type and mutant protein complexes. Experimental determination of $\Delta\Delta G$ values, while accurate, can be tedious and costly. Consequently, there has been a surge in the development of computational methods to predict these energy changes.
Broadly, these computation approaches fall into two main categories: physics-based and machine learning-based methods. The former, rooted in biophysical principles, delves into protein conformations and offers a rigorous approach\cite{gapsys2016accurate, kellogg2011role, bender2016protocols}. However, they often demand significant computational resources and are not always scalable. On the other hand, machine learning-based methods have gained popularity due to their scalability and rapid prediction capabilities. Leveraging the wealth of data from PPI datasets such as ASEdb\cite{thorn2001asedb}, PINT\cite{kumar2006pint}, ProTherm\cite{kumar2006protherm}, SKEMPI\cite{moal2012skempi,jankauskaite2019skempi}, and others\cite{sirin2016ab,geng2016exploring,jemimah2017proximate}, machine learning models like mCSM\cite{pires2014mcsm}, BindProf\cite{brender2015predicting}, iSEE\cite{geng2019isee}, MutaBind\cite{zhang2020mutabind2}, and several others\cite{jemimah2020proaffimuseq, zhou2020mutation, strokach2021elaspic2, liu2021deep} have been developed. These models have shown significant potential in predicting $\Delta\Delta G$s. However, challenges such as imbalanced training datasets, generalization across different PPI datasets, and the intricacy of capturing complex sequence-structure-function relationships remain obstacles \cite{fariselli2015inps, thiltgen2012assessing, pucci2018quantification, usmanova2018self}. This underscores the need for further research to enhance machine learning methodologies, ensuring accurate and efficient $\Delta\Delta G$ predictions.

In recent years, geometric graph learning has emerged as a promising approach for analyzing complex biomolecular systems\cite{nguyen2017rigidity,rana2023geometric,jiang2021ggl}. By representing proteins and their interactions as graphs, this methodology leverages the power of graph theory and machine learning to capture essential structural and spatial features of the biomolecular complexes. Specifically, the use of geometric subgraphs, which encode local interactions between atoms and residues, offers a rich representation. This not only sheds light on intricate molecular details but also provides insights into their impact on binding affinity\cite{rana2023geometric}. 

\begin{figure}[t]
    \centering
    \includegraphics[width=0.99\textwidth]{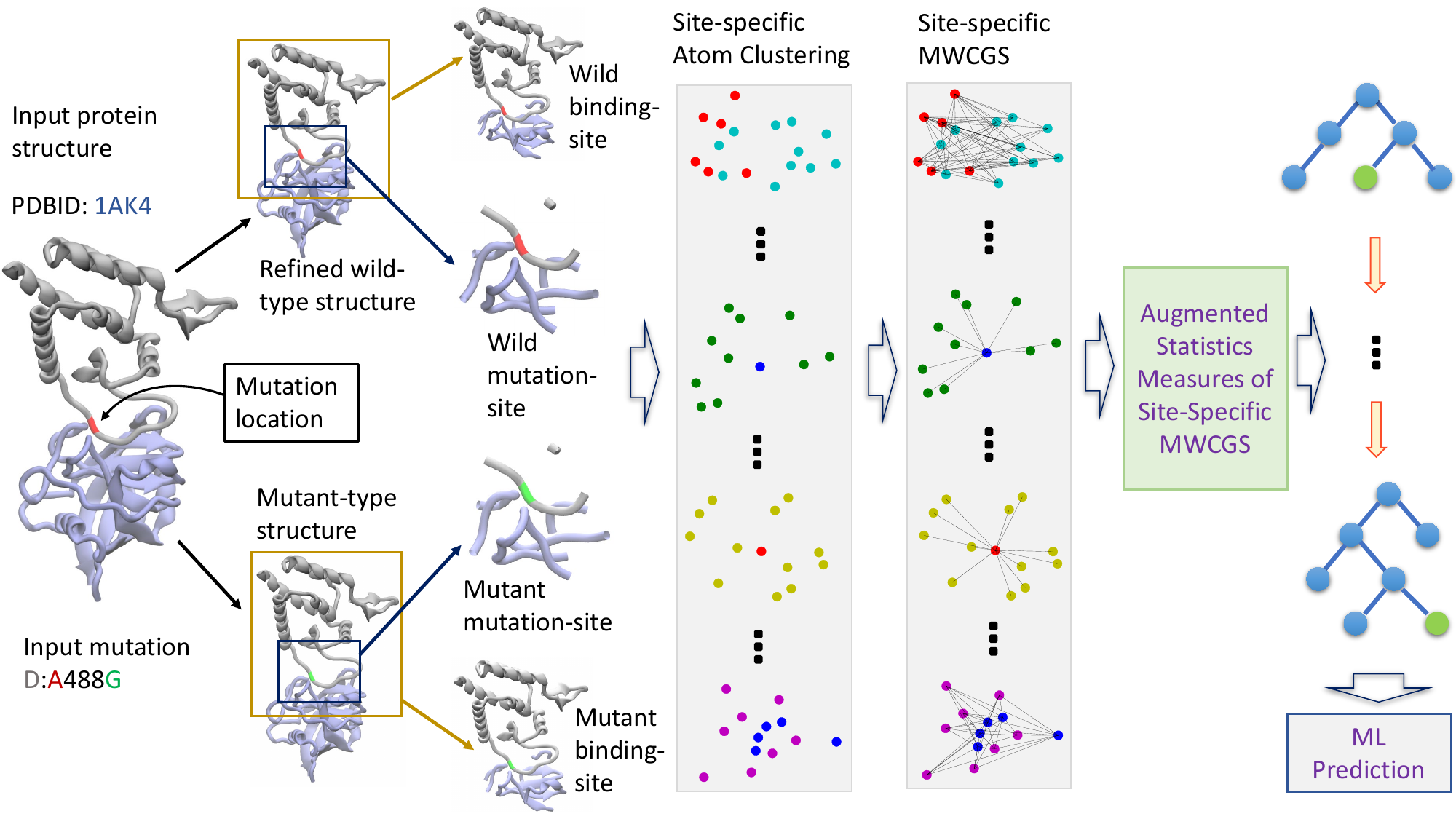}
    \caption{Illustration of the Geometric Graph Learning for Protein-Protein Interactions (GGL-PPI) workflow. Beginning on the left, an example protein structure (PDBID 1AK4) with a specific mutation  (D:A488G) is introduced. The central columns display the refined wild-type and mutant-type structures, processed using the JACKAL software, followed by the depiction of binding and mutation sites. Subsequent stages, detailed in columns four and five, involve the generation of Multi-Scale Weighted Colored Geometric Subgraphs (MWCGS) to capture geometric characteristics vital for protein interactions. The sixth column emphasizes feature augmentation, integrating statistical data on the rigidity of MWCGS at specific sites. Concluding on the right, the augmented features serve as input for ensemble learning methods, highlighted by gradient boosting trees. Further details are explored in Section \ref{sec:methods}.}
    \label{fig:GGL_PPI_Wrokflow}
\end{figure}

This work presents a novel method, called GGL-PPI (Geometric Graph Learning for Protein-Protein Interactions), which combines the principles of geometric graph learning and machine learning to predict mutation-induced binding free energy changes. The workflow of GGL-PPI is depicted in Figure \ref{fig:GGL_PPI_Wrokflow}. Central to its methodology, GGL-PPI utilizes atom-level graph coloring and multi-scale weighted colored geometric subgraphs, enabling the extraction of informative features from protein structures and their interactions. These features serve as inputs to a gradient-boosting tree model, which facilitates precise and consistent predictions of binding free energy change upon mutations. When compared with existing models, GGL-PPI consistently outperforms state-of-the-art approaches across all datasets. Further addressing its generalizability, GGL-PPI was evaluated on a blind test set, S$^{\mathrm{sym}}$ dataset\cite{pucci2018quantification}. This evaluation was conducted using a homology-reduced balanced training set to avert data leakage, showcasing GGL-PPI's robust performance and ability to produce unbiased predictions for both direct and reverse mutations.

\section{Datasets and Results}
In this section, we perform validation and evaluation of our proposed models on several benchmark datasets.  
We develop two types of GGL-PPI models: GGL-PPI1 and GGL-PPI2. The first model, GGL-PPI1, is built solely on geometric graph features discussed in Section \ref{sec:methods}. On the other hand, GGL-PPI2 incorporates both geometric graph features and auxiliary features, as detailed by Wang et al.\cite{wang2020topology}. The electrostatic potential calculations for the auxiliary components are conducted using the MIBPB software \cite{chen2011mibpb}.

\subsection{Validation}
To validate our models, we primarily consider the AB-Bind dataset \cite{sirin2016ab}, SKEMPI 1.0 dataset \cite{moal2012skempi}, and SKEMPI 2.0 dataset \cite{jankauskaite2019skempi}.
We employ a rigorous evaluation methodology by conducting a 10-times 10-fold cross-validation (CV) on each datasets. 
The mean Pearson correlation coefficient ($R_p$) and root-mean-square error (RMSE) serve as our evaluation metrics.
In comparing the CV performance of our proposed models with other existing methods, we specifically assess TopNetTree \cite{wang2020topology}, Hom-ML-V2\cite{liu2022hom}, and Hom-ML-V1 \cite{liu2022hom}. Both TopNetTree and Hom-ML-V2 incorporate auxiliary features in conjunction with their topology-based and Hom-complex-based features, respectively. On the other hand, Hom-ML-V1 solely relies on Hom-complex-based features without utilizing any auxiliary features.

\paragraph{Validation on AB-Bind S645 Data Set} The AB-Bind dataset contains 1,101 mutational data points for 32 antibody-antigen complexes, providing experimentally determined binding affinity changes upon mutations. Pires et al. curated a subset known as AB-Bind S645 \cite{pires2016mcsm}, consisting of 645 single-point mutations observed in 29 antibody-antigen complexes. The dataset comprises a mix of stabilizing (20\%) and destabilizing (80\%) mutations. 
Additionally, the dataset includes 27 non-binders that do not show any binding within the assay's sensitivity range. For these non-binders, the binding free energy changes have been uniformly set to a value of 8 kcal/mol.
It is crucial to consider these non-binders as outliers during model development and evaluation to ensure model accuracy and robustness.

\begin{figure}[t]
    \centering
    \includegraphics[width=\linewidth]{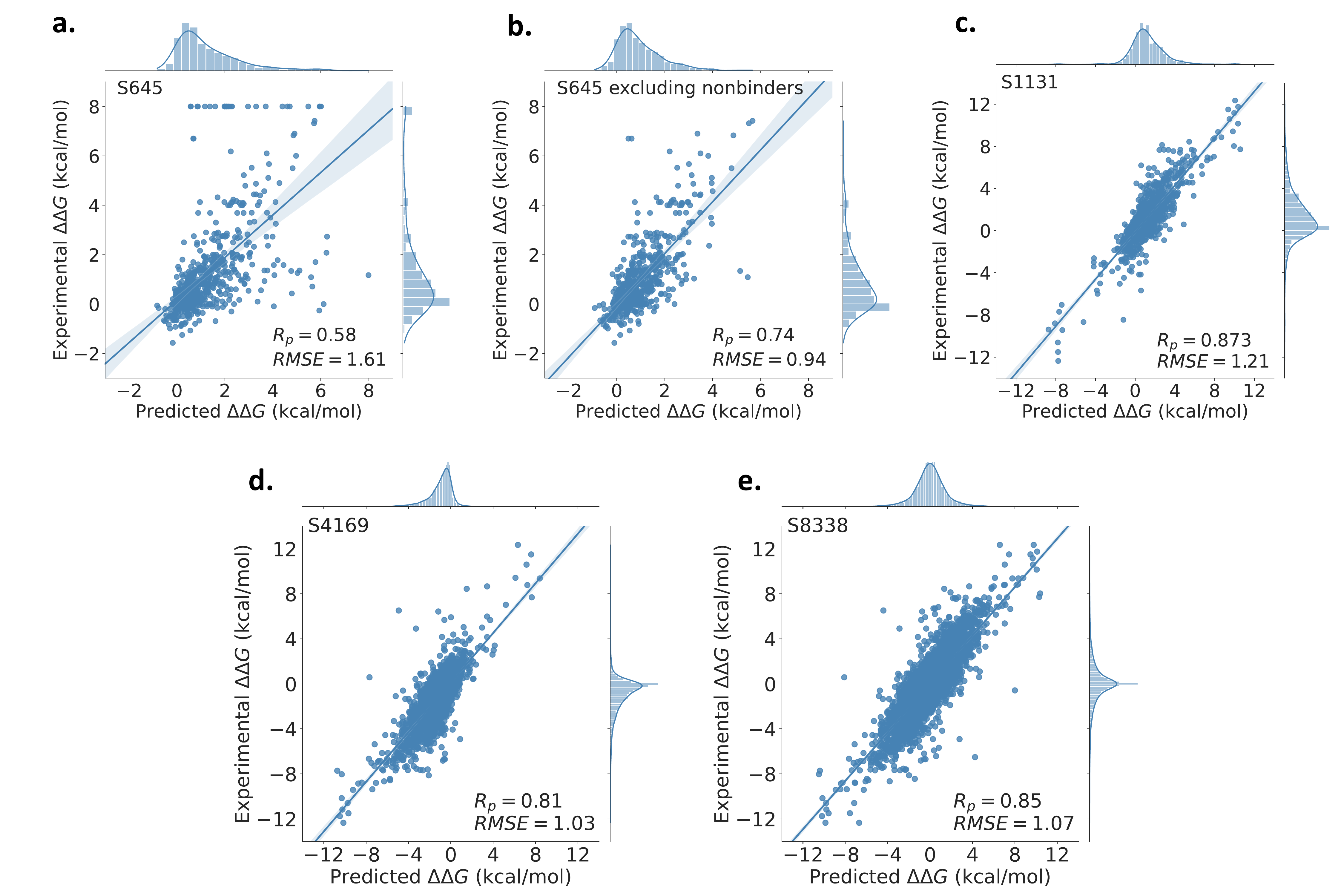}
    \caption{Performance of our GGL-PPI2 model on various validation datasets using 10-times 10-fold cross-validation. 
    (a) On the AB-Bind S645 dataset, our model achieves a Pearson's correlation coefficient ($R_p$) of 0.58 and a Root Mean Square Error (RMSE) of 1.61 kcal/mol. 
    (b) On the S645 dataset, excluding the 27 nonbinders, our model achieves an $R_p$ of 0.74 and an RMSE of 0.94 kcal/mol. 
    (c) On the SKEMPI 1.0 S1131 dataset, our model achieves an $R_p$ of 0.873 and an RMSE of 1.21 kcal/mol. 
    (d) On the SKEMPI 2.0 S4169 dataset, our model achieves an $R_p$ of 0.81 and an RMSE of 1.03 kcal/mol. 
    (e) On the S8338 dataset, our model achieves an $R_p$ of 0.85 and an RMSE of 1.07 kcal/mol.}
    \label{fig:kfold_results}
\end{figure}

Our GGL-PPI2 achieved an $R_p$ of 0.58 on the AB-Bind S645 dataset, as shown in Figure \ref{fig:kfold_results}a. The comparison results in Table \ref{tab:ab_bind_results} indicate that our model tied for second place with Hom-ML-V2 \cite{liu2022hom}, while TopNetTree \cite{wang2020topology} claimed the top position. However, when we exclude the 27 nonbinders from the dataset, our model outperforms all other existing models. Specifically, the $R_p$ value increases to 0.74 from 0.58 after removing the nonbinders (Figure \ref{fig:kfold_results}b).

Furthermore, GGL-PI1, our purely geometric graph-based features model, demonstrated competitive performance with an $R_p$ of 0.57 on the AB-Bind S645 dataset. Intriguingly, when excluding the nonbinders, GGL-PPI1 surpassed all other models with an improved $R_p$ of 0.73.  These performances reveal that our multiscale weighted colored geometric graphs can effectively characterize the wide range of interactions in biomolecular complexes.

\begin{table}[htbp]
    \centering
    \begin{tabular}{p{1cm}p{4cm}p{4cm}p{3cm}}
    \Tstrut\\
    \hline
    & & \multicolumn{2}{c}{$R_p$} \Tstrut\Bstrut\\
       &Method  &  with nonbinders & without nonbinders\Tstrut\Bstrut\\
   \hline
        &TopNetTree & 0.65 &0.68 \Tstrut\Bstrut\\
        &GGL-PPI2 & \textbf{0.58} &\textbf{0.74} \Bstrut\\
        &Hom-ML-V2 & 0.58 &0.70 \Bstrut \\
        &Hom-ML-V1 & 0.58 &0.68 \Bstrut \\
        &GGL-PPI1 & \textbf{0.57} &\textbf{0.73} \Bstrut\\
        &mCSM-AB & 0.53 &0.56\Bstrut \\
        &Discovery Studio & 0.45 &\Bstrut \\
        &mCSM-PPI & 0.35 &\Bstrut \\
        &FoldX &0.34 &\Bstrut\\
        & STATIUM & 0.32 & \Bstrut\\
        &DFIRE &0.31 & \Bstrut\\
        &bAsA &0.22 & \Bstrut\\
        &dDFIRE &0.19 & \Bstrut\\
        &Rosetta &0.16 & \Bstrut\\
    \hline
    \end{tabular}
    \caption{Performance comparison of different methods in terms of Pearson correlation coefficients ($R_p$) for the AB-Bind (S645) dataset.}
    \label{tab:ab_bind_results}
\end{table}

\paragraph{Validation on SKEMPI 1.0 S1131 Data Set} The SKEMPI 1.0 dataset consists of a collection of 3,047 mutations of 158 complexes obtained from literature sources, where the complexes have experimentally determined structures \cite{moal2012skempi}. The dataset includes both single-point mutations and multi-point mutations. Specifically, there are 2,317 entries in the dataset that represent single-point mutations, which are collectively known as the SKEMPI S2317 set. Additionally, a subset of 1,131 non-redundant interface single-point mutations has been selected from the SKEMPI S2317 set and labeled as the SKEMPI S1131 set \cite{xiong2017bindprofx}. This subset focuses on studying the impact of single-point mutations on protein-protein interactions.  

Figure \ref{fig:kfold_results}c shows that our model GGL-PPI2 achieves an $R_p$ of 0.873 and an RMSE of 1.21 kcal/mol in 10-fold CV on the S1131 dataset. Table \ref{tab:SKEMPI1.0_results} presents the performance comparison of various methods on the S1131 dataset, including our proposed models, GGL-PPI1 and GGL-PPI2. Among them, our model, GGL-PPI2, achieved the highest performance, underscoring its superiority in predicting binding affinity changes due to mutation. Notably, even without auxiliary features, our GGL-PPI1 outperformed both TopNetTree and Hom-ML-V2 methods that do leverage auxiliary features. This again highlights the efficacy of our geometric graph-based molecular representation.

\begin{table}[htbp]
    \centering
    \begin{tabular}{p{1cm}p{6cm}p{2cm}}
    \Tstrut\\
    \hline
       &Method  &  $R_p$ \Tstrut\Bstrut\\
   \hline
        &GGL-PPI2 & \textbf{0.873} \Tstrut\Bstrut\\
        &GGL-PPI1 & \textbf{0.865} \Bstrut\\
        &Hom-ML-V2 & 0.857 \Bstrut\\
        &TopNetTree & 0.850 \Bstrut\\
        &Hom-ML-V1 & 0.792 \Bstrut\\
        &BindProfX & 0.738 \Bstrut\\
        &Profile-score+FoldX &0.738 \Bstrut\\
        &Profile-score & 0.675 \Bstrut\\
        &SAAMBE & 0.624 \Bstrut\\
        &FoldX & 0.457 \Bstrut\\
        &BeAtMuSic & 0.272 \Bstrut\\
        &Dcomplex & 0.056 \Bstrut\\    
    \hline
    \end{tabular}
    \caption{Performance comparison of different methods in terms of Pearson correlation coefficients ($R_p$) for the single-point mutations in the SKEMPI 1.0 (S1131) dataset.}
    \label{tab:SKEMPI1.0_results}
\end{table}

\paragraph{Validation on SKEMPI 2.0 S4169 and S8338 Data Sets} The SKEMPI 2.0 dataset is an updated and expanded version of the original SKEMPI dataset, incorporating new mutations collected from various sources \cite{jankauskaite2019skempi}. Released in 2018, it significantly increased in size, now containing a total of 7,085 entries, including both single-point and multi-point mutations. The data was obtained by merging several databases, including SKEMPI 1.0 \cite{moal2012skempi}, AB-Bind\cite{sirin2016ab}, PROXiMATE \cite{jemimah2017proximate}, and dbMPIKT \cite{liu2018dbmpikt}. Additionally, new data from the literature were manually curated and added to the dataset. The mutations cover a wide range of protein complexes, such as protease-inhibitor, antibody-antigen, and TRC-pMHC complexes. Among the mutations, approximately 3,000 are single-point alanine mutations, 2,000 are single-point non-alanine mutations, and another 2,000 involve multiple mutations.

Notably, the authors of the mCSM-PPI2 \cite{rodrigues2019mcsm} method filtered the single-point mutations, yielding S4169 set, comprising 4,169 variants in 139 different complexes/
The S8338 set, derived from S4169, represents hypothetical reverse mutation energy changes with negative values. This comprehensive dataset serves as a valuable resource for studying protein interactions and their thermodynamic properties.

Perforamnce-wise, Our GGL-PPI2 model posts an $R_p$ of 0.81 with an RMSE of 1.03 kcal/mol for the S4169 dataset as shown in Figure \ref{fig:kfold_results}d, outstripping all existing models (Table \ref{tab:SKEMPI2.0_results}). It is noteworthy that our GGL-PPI1 model, which solely relies on geometric graph-based features, demonstrated comparable performance to GGL-PPI2, outperforming TopNetTree and mCSM-PPI2 with an $R_p$ of 0.80 and an RMSE of 1.06 kcal/mol.

In the case of the S8338 dataset, we applied a stratified cross-validation approach similar to mCSM-PPI2. We ensured that hypothetical reverse mutations were consistently placed either in the training or test sets during the dataset splits, maintaining their relationship to the corresponding original mutations intact throughout the cross-validation process.
GGL-PPI2 achieved an $R_p$ of 0.85 with an RMSE of 1.07 kcal/mol as depicted in Figure \ref{fig:kfold_results}e, and GGL-PPI1 closely followed, attaining 
an $R_p$ of 0.84 with the same RMSE value. As Table \ref{tab:SKEMPI2.0_results} attests,
our GGL-PPI2 is on par with TopNetTree and outperforms mCSM-PPI2 on the S8338 dataset.


\begin{table}[htbp]
    \centering
    \begin{tabular}{p{1cm}p{6cm}p{4cm}p{2cm}}
    \Tstrut\\
    \hline
    & & \multicolumn{2}{c}{$R_p$} \Tstrut\Bstrut\\
       &Method  &  S4169 & S8338 \Tstrut\Bstrut\\
   \hline
        &GGL-PPI2 & \textbf{0.81} & \textbf{0.85}\Tstrut\Bstrut\\
        &GGL-PPI1 & \textbf{0.80} & \textbf{0.84}\Tstrut\Bstrut\\
        &Hom-ML-V2 & 0.80 &-- \Bstrut\\
        &TopNetTree & 0.79 &0.85 \Bstrut\\
        &Hom-ML-V1 & 0.77 &-- \Bstrut\\
        &mCSM-PPI2 & 0.76 &0.82\Bstrut\\  
    \hline
    \end{tabular}
    \caption{Performance comparison of different methods in terms of Pearson correlation coefficients ($R_p$) for the single-point mutations in the SKEMPI 2.0 (S4169 and S8338) dataset.}
    \label{tab:SKEMPI2.0_results}
\end{table}

\subsection{Evaluation}
To evaluate our proposed model for predicting binding free energy (BFE) changes of protein-protein interactions, we consider two datasets sourced from the ProTherm database \cite{kumar2006protherm}.

The first dataset, carefully selected by Pucci et al. \cite{pucci2018quantification}, named S$^{\mathrm{sym}}$ dataset. This data assembles 684 mutations from the ProTherm, comprising 342 direct mutations and their corresponding reverse mutations, resulting in a balanced dataset. The dataset specifically focuses on mutations in fifteen protein chains with solved 3D structures, ensuring high-resolution data with a resolution of at least 2.5\AA. By providing experimentally measured $\Delta\Delta G$ values and a balanced representation of stabilizing and destabilizing mutations, the S$^{\mathrm{sym}}$ dataset serves as a valuable resource for evaluating prediction biases in the context of predicting mutation-induced binding affinity changes.

To address the issue of data leakage and enhance the generalization capability of our method, we employed the Q1744 dataset \cite{li2020predicting}. Quan et al. \cite{quan2016strum} compiled the Q3421 dataset from ProTherm, consisting of 3421 single-point mutations across 150 proteins with available PDB structures. However, the presence of homologous proteins in both the training and test set can lead to interdependent effects of mutations, compromising the model's performance. To mitigate this, Li et al. \cite{li2020predicting} created the Q1744 dataset, derived by excluding overlapping data points and refining protein-level homology between Q3421 and S$^{\mathrm{sym}}$ datasets, resulting in 1744 distinct mutations.
Furthermore, the Q3488 dataset was created by augmenting reverse mutations in the Q1744 set. 
We utilized the Q3488 dataset as our training set, thereby enhancing our $\Delta\Delta G$ predictor's capability to accurately predict BFE changes in PPIs.

We conduct an evaluation of our model on the blind test set S$^{\mathrm{sym}}$,  with a distinct focus on both direct and reverse mutations. To assess the performance, we utilize the Pearson correlation coefficient and root-mean-square error as our primary metrics. Additionally, to discern any prediction bias, we incorporated two statistical measures: $R_{p_{\mathrm{dir-rev}}}$ and $\delta$. The former calculates the Pearson correlation between predictions for direct and reverse mutations, while the latter represents the sum of predicted $\Delta \Delta G$ values for both types of mutations. The hypothesis is that an unbiased predictor would yield $R_{p_{\mathrm{dir-rev}}}=-1$ and an average $\delta$ ($\bar\delta$) of 0 kcal/mol.

\begin{figure}[t]
    \centering
    \includegraphics[width=\linewidth]{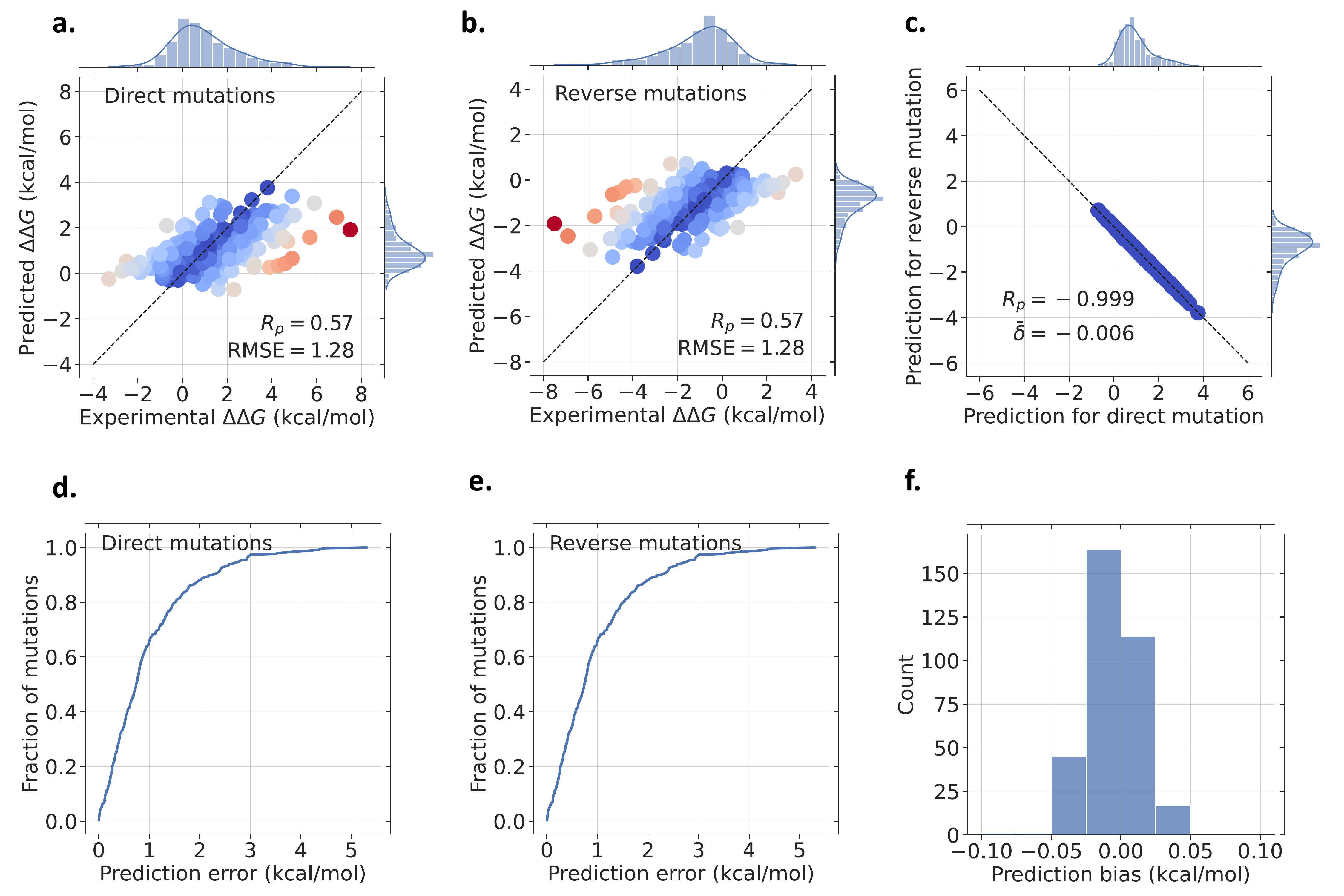}
    \caption{Results of our GGL-PPI2 model for S$^{\mathrm{sym}}$ dataset. In (a), direct mutations are plotted, while (b) presents the results for reverse mutations. The color spectrum, ranging from blue to red, represents the corresponding prediction accuracy—where blue signifies higher accuracy and red indicates lower accuracy. A comparison between direct and reverse mutations is illustrated in (c). Cumulative error distributions for direct and reverse mutations are displayed in (d) and (e), respectively. The prediction bias is visualized in (f) through a histogram plot.}
    \label{fig:s_sym_results}
\end{figure}

Our main focus is to highlight the effectiveness of our model, GGL-PPI2,  particularly emphasizing its robust geometric graph-based molecular featurization. GGL-PPI2 has demonstrated exceptional prediction accuracy, maintaining consistency for both direct and reverse mutations. As depicted in Figure \ref{fig:s_sym_results}a and \ref{fig:s_sym_results}b, our model achieves consistent $R_p$ values of 0.57 and an RMSE of 1.28 kcal/mol, indicating its efficiency against overfitting to direct mutations.

Additionally, the analysis reveals that a significant proportion of mutations fall within a prediction error of 0.5 kcal/mol and 1.0 kcal/mol, with 34.6\% and 65.8\% for direct mutations and 35.1\% and 66.0\% for reverse mutations, as depicted in Figure \ref{fig:s_sym_results}d and \ref{fig:s_sym_results}e. Furthermore, Figure \ref{fig:s_sym_results}c demonstrates that GGL-PPI2 effectively addresses prediction bias by achieving a nearly perfect $R_{p_{\mathrm{dir-rev}}}$ value of -0.999 and an extremely low average $\bar\delta$ of 0.006 kcal/mol. Finally, the distribution plot in Figure \ref{fig:s_sym_results}f illustrates that 99.4\% of mutations exhibit a prediction bias under 0.05 kcal/mol. 

In Table \ref{tab:s_sym_results}, we present the prediction results of our models and conduct a comprehensive comparison with other $\Delta\Delta G$ predictors. We observe that our GGL-PPI2 model outperforms ThermoNet\cite{li2020predicting}, which was also trained on the homology-reduced set Q3488, across all evaluation measures. It outperforms ThermoNet by 21.3\% for direct mutations and 18.7\% for reverse mutations. Furthermore, the GGL-PPI1 model, which only uses geometric graph-based features, also performs better than ThermoNet in both direct and reverse prediction tasks. This further emphasizes the effectiveness of our geometric-graph approach.

For a broader comparison against other $\Delta\Delta G$ predictors, we introduce the GGL-PPI2$^{\ast}$ model, trained on the Q6428 set constructed before the homology reduction of the set Q3421 \cite{li2020predicting}. As illustrated in Table \ref{tab:s_sym_results}, GGL-PPI2$^{\ast}$ excels over other methods in reverse mutation predictions. It is noteworthy that while some methods surpass GGL-PPI2$^{\ast}$ for direct mutations, they frequently exhibit significant bias towards reverse mutations.



\begin{table}[htbp]
\begin{threeparttable}
    \centering
    \begin{tabular}{p{3cm}p{2cm}p{2cm}p{2cm}p{2cm}p{2cm}p{1cm}}
    \Tstrut\\
    \hline
      Method\tnote{b} & RMSE$_{\mathrm{dir}}$ & $R_{p_{\mathrm{dir}}}$ & RMSE$_{\mathrm{rev}}$ & $R_{p_{\mathrm{rev}}}$ & $R_{p_{\mathrm{dir-rev}}}$ &$\bar\delta$ \Tstrut\Bstrut\\
    \hline
      GGL-PPI2$^{\ast}$ & 1.22 &0.66 &1.22 &0.66 &-0.99 &0.0003\Tstrut\Bstrut\\
      GGL-PPI1$^{\ast}$ & 1.34 &0.61 &1.34 &0.61 &-0.99 &-0.01\Tstrut\Bstrut\\
      ThermoNet$^{\ast}$ & 1.42 &0.58 &1.38 &0.59 &-0.95 &-0.05\Tstrut\Bstrut\\
      GGL-PPI2 & \textbf{1.28} &\textbf{0.57} &\textbf{1.28} &\textbf{0.57} &\textbf{-0.99} &\textbf{0.006}\Tstrut\Bstrut\\
      GGL-PPI1 & \textbf{1.32} & \textbf{0.53} &\textbf{1.32} &\textbf{0.53} &\textbf{-0.99} &\textbf{0.004}\Tstrut\Bstrut\\
      DDGun3D &1.42 &0.56 &1.46 &0.53 &-0.99 &-0.02\Bstrut\\
      DDGun &1.47 &0.48 &1.50 &0.48 &-0.99 &-0.01\Bstrut\\
      ThermoNet & 1.56 &0.47 &1.55 &0.48 &-0.96 &-0.01\Bstrut\\
      PoPMuSiC$^{\mathrm{sym}}$ &1.58 &0.48 &1.62 &0.48 &-0.77 &0.03\Bstrut\\
      MAESTRO &1.36 &0.52 &2.09 &0.32 &-0.34 &-0.58\Bstrut\\
      FoldX &1.56 &0.63 &2.13 &0.39 &-0.38 &-0.47\Bstrut\\
      PoPMuSiC 2.1 &1.21 &0.63 &2.18 &0.25 &-0.29 &-0.71\Bstrut\\
      SDM &1.74 &0.51 &2.28 &0.32 &-0.75 &-0.32\Bstrut\\
      iSTABLE &1.10 &0.72 &2.28 &-0.08 &-0.05 &-0.60\Bstrut\\
      I-Mutant 3.0 &1.23 &0.62 &2.32 &-0.04 &0.02 &-0.68\Bstrut\\
      NeEMO &1.08 &0.72 &2.35 &0.02 &0.09 &-0.60\Bstrut\\
      DUET &1.20 &0.63 &2.38 &0.13 &-0.21 &-0.84\Bstrut\\
      mCSM &1.23 &0.61 &2.43 &0.14 &-0.26 &-0.91\Bstrut\\
      MUPRO &0.94 &0.79 &2.51 &0.07 &-0.02 &-0.97\Bstrut\\
      STRUM &1.05 &075 &2.51 &-0.15 &0.34 &-0.87\Bstrut\\
      Rosetta &2.31 &0.69 &2.61 &0.43 &-0.41 &-0.69\Bstrut\\
      AUTOMUTE &1.07 &0.73 &2.61 &-0.01 &-0/06 &-0.99\Bstrut\\
      CUPSAT &1.71 &0.39 &2.88 &0.05 &-0.54 &-0.72\Bstrut\\
    \hline
    \end{tabular}
    \caption{Comparison of various methods for the balanced test set S$^{\mathrm{sym}}$.\tnote{a,c}}
    \label{tab:s_sym_results}
    \begin{tablenotes}
        \item[a] 
        RMSE and $R_p$ represent the root-mean-square error and Pearson correlation, respectively. The subscripts ``dir'' and ``rev'' indicate results for direct and reverse mutations, respectively.
        $R_{p_{\mathrm{dir-rev}}}$ represents the Pearson correlation between the predicted $\Delta\Delta G$s for direct and reverse mutations.
        $\bar\delta$ is the average of $\delta=\Delta\Delta G_{\mathrm{dir}} + \Delta\Delta G_{\mathrm{rev}}$, which represents the average prediction bias.
        \item[b] The asterisk ($\ast$) in the methods ThermoNet and GGL-PPI indicates that they were trained on the set Q6428, which was constructed before the homology reduction of the set Q3421.
        \item[c] Results of other methods except for ours are collected from Li et al. \cite{li2020predicting}.
    \end{tablenotes}
\end{threeparttable}
\end{table}

\section{Methods}\label{sec:methods}

\subsection{Graph Theory and Atom-level Interactions in Biomolecules}

Graph theory provides a mathematical framework that is widely applied in the study of biomolecules such as proteins, DNA, and RNA. For a biomolecule, a graph $G(\mathcal{V}, \mathcal{E})$ is a collection of nodes $\mathcal{V}$ and edges $\mathcal{E}$ that can represent the connectivity and relationships between different atoms or residues within the molecule.

A refinement to this representation is graph coloring, a technique that assigns unique labels to different atom types within the biomolecule. This enriched, colored graph encodes diverse atomic interactions, paving the way for a collective and coarse-grained description of the dataset. In this representation, atoms with assigned labels are organized into subgraphs, and the colored edges between them represent atom-specific interactions.

The advantage of using subgraphs lies in their ability to focus on specific regions or components of the biomolecule. By isolating relevant subsets of atoms, subgraphs allow us to identify localized patterns, interactions, or clusters that might not be evident in the global graph representation. This targeted approach provides a more nuanced understanding of the structural and functional properties of biomolecules.

To extract atom-level interaction information, we consider specific atom types based on their names in the PDB structure such as carbon alpha (CA), carbon beta (CB), carbon delta-1 (CD1), etc. These atom names serve as identifiers for specific positions within a protein's three-dimensional structure. They help define the individual atoms that constitute amino acids, the building blocks of proteins, and provide crucial information about their spatial orientation and chemical properties. We consider a total of 37 distinct atom names that are frequently found in protein structures within the PDB database. These atom types are represented by the set $\mathcal{A}$. 
 
To simplify the notation, we assume the set $\mathcal{A}$ is sorted in alphanumeric order, 
\begin{equation}
    \mathcal{A} = \{\mathrm{C, CA, CB, \cdots, N, ND1, ND2, \cdots, O, OD1, \cdots, SD, SG} \},
\end{equation}
and $\mathcal{A}_k$ represents the $k$th element of the set, e.g. $\mathcal{A}_0=C,\,\mathcal{A}_1=CA$, etc.  
This extended atom-level graph coloring scheme has been shown to demonstrate superior performance in predicting protein-ligand binding affinity, as demonstrated in our previous work \cite{rana2023geometric}.

By utilizing this comprehensive set of atom types, we can construct a weighted colored subgraph that captures the intricate relationships between different atoms in a biomolecular system. The subgraph's vertices, denoted by $\mathcal{V}$, are defined by the coordinates $\mathbf{r}_i$ of each atom, along with its associated atom type $\alpha_i$. Formally, $\mathcal{V}$ can be expressed as:
\begin{equation}
    \mathcal{V}=\{ (\mathbf{r}_i,\alpha_i)| \mathbf{r}_i\in \mathbb{R}^3; \alpha_i\in \mathcal{A}; i=1,2,\cdots,N\}.
\end{equation}
To define the edges $\mathcal{E}$ of the subgraph, we consider the characteristic distance $\eta_{kk'}$ between pairs of atom types $\mathcal{A}_k$ and $\mathcal{A}_{k'}$. We use a subgraph weight function $\Phi$ to determine the weight of each edge. The edges $\mathcal{E}$ can be defined as follows:
\begin{align}
    \mathcal{E} &= \{ \Phi(\norm{\mathbf{r}_i-\mathbf{r}_j};\eta_{kk'})|  \alpha_i=\mathcal{A}_k,\, \alpha_j = \mathcal{A}_{k'};\, i,j=1,2,\cdots,N \}.
\end{align}
Here, $\norm{\mathbf{r}_i-\mathbf{r}_j}$ represents the Euclidean distance between the $i$th and $j$th atoms. The weight function $\Phi$ quantifies the strength of interaction between atoms based on their Euclidean distance. A commonly used choice for $\Phi$ is the generalized exponential function or the generalized Lorentz function. For instance, the generalized exponential function is defined as:
\begin{equation}
    \Phi_E(\norm{\mathbf{r}_i-\mathbf{r}_j};\eta_{kk'}) = e^{-(\norm{\mathbf{r}_i-\mathbf{r}_j}/\eta_{kk'})^\kappa}, \quad \kappa>0,
\end{equation}
The resulting weighted colored subgraph $G(\mathcal{V}, \mathcal{E})$ provides a powerful representation of the molecular properties at the atomic level. By analyzing the subgraph, we can extract collective molecular descriptors and investigate the multiscale behavior of the system. This multiscale behavior arises from considering different characteristic distances $\eta_{kk'}$ for various pairs of atom types, enabling the generation of a wide range of scalable graph-based descriptors. The geometric subgraph centrality, defined as 
\begin{align}\label{MWCGS_rigidity}
     \mu^G(\eta_{kk'}) &=\sum_i \mu_i^G(\eta_{kk'})=\sum_i \sum_j \Phi(\norm{\mathbf{r}_i-\mathbf{r}_j};\eta_{kk'}),\nonumber \\
     &\quad \alpha_i = \mathcal{T}_{k},\, \alpha_j = \mathcal{T}_{k'},
 \end{align}
serves as a measure of the combined strength of interaction between chosen pairs of atom types, providing valuable insights into the molecular structure and properties.

\subsection{Geometric Subgraph Representation of PPIs}


In the context of studying protein-protein interactions (PPIs) and predicting the effects of mutations on these interactions, it is important to focus on the relevant regions where the interactions occur. While protein-protein complexes can consist of a large number of atoms, the interactions between proteins primarily take place at specific regions known as interfaces. To streamline computational costs and concentrate on pertinent information, it is common practice to consider only the protein atoms near the binding sites.

The binding site, in this context, refers to the region within a certain cutoff distance $c$ from the chain where the mutation occurred. By defining the binding site in this way, we can narrow our focus to the specific area where the interaction and subsequent effects of the mutation are most pronounced. Furthermore, when analyzing the effects of mutations, it is crucial to incorporate geometric graph information from the mutation sites and their neighboring regions. The mutation site is defined as the region within a cutoff distance $c$ from the mutated residue, allowing us to capture the structural changes resulting from the mutation.

To construct a site-specific multiscale weighted colored geometric subgraph (MWCGS) representation for a PPI, both the wild-type and mutant-type proteins are considered. This leads to four sets of features for each PPI, corresponding to the two sites and the two types of proteins involved. Each set consists of $37\times 37=1369$ MWCGS features, representing the interactions between the atom types involved in the PPI. 
These features encompass diverse chemical and biological properties, such as the presence of specific interatomic interactions involving oxygen and nitrogen atoms, the hydrophobic nature of certain regions, and the ability of atoms to undergo polarization, among other relevant molecular characteristics. By utilizing these site-specific MWCGS features, we can uncover valuable insights into the effects of mutations and the underlying molecular interactions, revealing significant information and characteristics embedded within the PPI system.

\subsection{Geometric Graph Learning for PPIs}
Accurately predicting the changes in binding affinity induced by mutations in protein-protein complexes poses a significant challenge due to the complex nature of these systems. The interactions between proteins are highly intricate, and the effects of mutations can be subtle and context-dependent. Machine learning techniques offer a promising approach to tackle this problem by leveraging the power of data-driven models to capture complex patterns and relationships.

Machine learning algorithms can aid in predicting mutation-induced binding affinity changes by learning from a set of training examples that consist of protein-protein complexes with known experimental binding affinities. These algorithms can analyze the features extracted from the complexes, such as geometric graph information, to identify relevant patterns and associations between the features and the binding affinities. By learning from these patterns, the algorithms can generalize and make predictions on unseen protein-protein complexes.

There are several machine learning algorithms that can be used in combination with geometric graph features to predict binding affinity changes. These algorithms include random forests \cite{breiman2001random}, support vector machines (SVM) \cite{cortes1995support}, neural networks \cite{nguyen2020mathdl}, and gradient boosting trees (GBT) \cite{friedman2001greedy}. Each algorithm has its strengths and weaknesses, and their performance can vary depending on the specific problem and dataset.

Among these algorithms, gradient boosting trees (GBT) have gained significant popularity in recent years \cite{rana2023geometric}. GBT is an ensemble method that builds a sequence of weak learners, typically decision trees, to correct the errors made by the previous learners. By combining these weak learners, GBT can effectively model complex relationships and improve prediction accuracy. One advantage of GBT is its robustness against overfitting, which is especially beneficial when dealing with a moderate number of features. Additionally, GBT models can provide interpretability, allowing us to gain insights into the factors contributing to the binding affinity changes.

The implementation of the GBT algorithm in this study utilized the scikit-learn package (v 0.24.1). To optimize the performance of the GBT model for ensemble methods, specific hyperparameters were fine-tuned. The number of estimators was set to 40000, indicating the number of weak learners in the ensemble, while the learning rate was set to 0.001, determining the contribution of each weak learner to the final prediction.
Given the large number of features involved in the prediction task, an efficient training process was achieved by limiting the maximum number of features considered to the square root of the descriptor length. This approach helped expedite the training process without compromising the overall performance of the GBT model.
To ensure reliable performance evaluation, fifty runs were performed for each feature set, employing different random seeds. By averaging the results obtained from these runs, a more robust and representative performance measure was obtained. Despite the complexity of the prediction task and the involvement of numerous features, the selected parameter settings and multiple runs yielded satisfactory performance results.

The GBT approach was chosen for its ability to effectively handle overfitting, exhibit good performance with moderately sized datasets, and provide interpretable models. These characteristics make GBT a suitable and reliable choice for this study, enabling accurate predictions of mutation-induced binding affinity changes in protein-protein complexes using the provided geometric graph features.

\section{Conclusion}
The study of protein-protein interactions (PPIs) and the prediction of mutation-induced binding free energy changes are of great importance in understanding the molecular basis of biological processes. The application of geometric graph theory and atom-level graph coloring techniques provides a powerful framework for analyzing biomolecules and capturing their intricate relationships.
By utilizing the concept of geometric subgraphs and constructing multi-scale weighted colored geometric subgraphs (MWCGS), we can effectively represent the structural and functional properties of PPIs. The site-specific MWCGS features allow us to extract meaningful patterns and characteristics, shedding light on the effects of mutations and the underlying molecular interactions.

In this work, we developed a mutation-induced binding free energy change predictor, called GGL-PPI, by incorporating site-specific MWCGS features for PPIs and gradient-boosting trees. Our method demonstrates superior performance compared to existing methods. The model was validated on three datasets: AB-Bind S645, SKEMPI 1.0 S1131, and SKEMPI 2.0 S4169 and S8338, showcasing its robustness and effectiveness.
Furthermore, GGL-PPI was evaluated on a blind test set, the S$^{\mathrm{sym}}$ dataset. To prevent data leakage between the test and training sets, the model was trained on a homology-reduced balanced training set Q3488. This approach ensures the reliability and fairness of the evaluation process. GGL-PPI exhibits the most unbiased and superior performance in predicting binding free energy changes for both direct and reverse mutations, outperforming other existing methods, particularly for reverse mutations. 

Overall, the results highlight the potential of the GGL-PPI approach in accurately predicting mutation-induced binding free energy changes in protein-protein interactions, providing valuable insights into the molecular mechanisms underlying protein-protein interactions and facilitating drug design and discovery efforts.

\section{Data and Software Availability}
The source code is available at Github: \url{https://github.com/NguyenLabUKY/GGL-Mutation}.

\section{Competing interests}
No competing interest is declared.

\section{Acknowledgments}
This work is supported in part by funds from the National Science Foundation (NSF: \# 2053284, \# 2151802, and \# 2245903), and the University of Kentucky Startup Fund.

\bibliographystyle{unsrt}
\bibliography{refs}

\end{document}